\documentclass[11pt]{article}
\usepackage{bm}
\usepackage[normalem]{ulem}
\usepackage[utf8]{inputenc}
\usepackage[T1]{fontenc}
\usepackage{mathptmx}
\usepackage{etoolbox}
\usepackage{xcolor}
\usepackage{physics}
\usepackage{footmisc}
\usepackage[a4paper, margin=0.9in]{geometry}
\usepackage{authblk}
\usepackage{graphicx}
\usepackage{dcolumn}
\usepackage{amsmath}
\usepackage{amssymb}
\usepackage{textcomp}
\usepackage{verbatim}
\usepackage{float}

\usepackage{tabularx}
\usepackage{booktabs}
\usepackage{makecell}

\usepackage{hyperref}
\hypersetup{
	colorlinks=true,
	linkcolor=blue,
	filecolor=blue,
	urlcolor=blue,
	citecolor = blue,
	pdfauthor=author,
}
\usepackage{multirow}
\usepackage[T1]{fontenc}
\usepackage[normalem]{ulem}
\usepackage[numbers,sort&compress]{natbib}

\makeatletter
\def\@fnsymbol#1{%
	\ifcase#1\or $\dagger$\or $*$\or $\ddagger$\or $\S$\or $\P$\or $\|$\or $**$\or $\dagger\dagger$\or $\ddagger\ddagger$\else\@ctrerr\fi}
\makeatother

\begin{document}

\title{Towards Environmentally Responsive Hypersound Materials}

\author[1]{Edson R. Cardozo de Oliveira}
\author[2]{Gastón Grosman}
\author[1]{Chushuang Xiang}
\author[2]{Michael Zuarez-Chamba}
\author[2]{Priscila Vensaus}
\author[1]{Abdelmounaim Harouri}
\author[3]{Cédric Boissiere}
\author[2]{Galo J. A. A. Soler-Illia\footnote{gsoler-illia@unsam.edu.ar}}
\author[1]{Norberto Daniel Lanzillotti-Kimura\footnote{daniel.kimura@cnrs.fr}}

\affil[1]{Université Paris-Saclay, C.N.R.S., Centre de Nanosciences et de Nanotechnologies (C2N),  10 Boulevard Thomas Gobert, 91120 Palaiseau, France}

\affil[2]{Instituto de Nanosistemas, Escuela de Bio y Nanotecnología, Universidad Nacional de San Martín-CONICET, Av. 25 de Mayo 1169, San Martín, Buenos Aires, Argentina}

\affil[3]{Laboratoire de Chimie de la Matière Condensée, Université Pierre et Marie Curie, 4 Place Jussieu, 75252 Paris, Cedex 5, France}

\date{}

\maketitle

\begin{abstract}

The engineering of acoustic phonons in the gigahertz (GHz) range holds significant potential for technological breakthroughs in areas such as data processing, sensing and quantum communication. Novel approaches for nanophononic resonators responsive to external stimuli provide additional control and functionality for these devices. Mesoporous thin films (MTFs) for example, featuring nanoscale ordered pores, support GHz-range acoustic resonances. These materials are sensitive to environmental changes, such as liquid and vapor infiltration, modifying their effective optical and elastic properties. Here, a SiO$_{2}$ MTF-based open-cavity nanoacoustic resonator is presented, in which the MTF forms the topmost layer and is exposed to the environment. Using a transient reflectivity setup, acoustic responses under varying humidity conditions are investigated. A pronounced shift in acoustic resonance frequency with changes in relative humidity is observed for the first time, demonstrating a simple way to tune hypersound confinement. In addition, resonators with varying pore sizes and thicknesses are compared, revealing that resonance frequencies are primarily influenced by material properties and film thickness, rather than pore size. The proposed open-cavity resonator design provides a versatile platform for future studies on the mechanical response of MTFs to liquid and vapor infiltration, opening the gate to environment-responsive hypersound devices.

\end{abstract}

\section{Introduction}

The engineering and manipulation of acoustic waves at the nanoscale have led to significant advancements across multiple fields including optoelectronics,\cite{stillerCoherentlyRefreshingHypersonic2020,daineseStimulatedBrillouinScattering2006} imaging and nondestructive testing,\cite{lomonosovNanoscaleNoncontactSubsurface2012,mechriDepthprofilingElasticInhomogeneities2009,sandeepInSituImagingLightInduced2022,wangImagingGrainMicrostructure2020} quantum technologies,\cite{chafatinosPolaritondrivenPhononLaser2020} and the demonstration of fundamental wave phenomena.\cite{arreguiAndersonPhotonPhononColocalization2019,lanzillotti-kimuraBlochOscillationsTHz2010,ortizTopologicalOpticalPhononic2021,ortizPhononEngineeringSuperlattices2019,arreguiCoherentGenerationDetection2019,rodriguezTopologicalNanophononicInterface2023} Recent progress in nanophononics has triggered the interest in developing novel functionalities, particularly in the domain of environment-responsive tunable nanophononic resonators.\cite{priyaPerspectivesHighfrequencyNanomechanics2023} Such functional resonators would drive advancements in sensing applications, as well as on the concept of phononic networks, unlocking new capabilities of nanoacoustic devices.

Mesoporous thin films (MTFs), characterized by self-assembled nanoscale-ordered pores, appear as logical candidates for complex nanophononic applications. These materials, produced using low-cost, soft fabrication techniques, support acoustic resonances in the 5-100 gigahertz (GHz) frequency range.\cite{abdalaMesoporousThinFilms2020,cardozodeoliveiraProbingGigahertzCoherent2023} Their high surface area allows for chemical functionalization of the nanopores, making them responsive to external stimuli.\cite{albertiGatedSupramolecularChemistry2015,calvoMesoporousAminopropylFunctionalizedHybrid2008,pizarroDropletsUnderlyingChemical2022a} In addition, MTFs are also susceptible to liquid and vapor infiltration into the pores, altering their optical and elastic properties,\cite{benettiPhotoacousticSensingTrapped2018,boissierePorosityMechanicalProperties2005,cardozodeoliveiraDesignCosteffectiveEnvironmentresponsive2023,gorElasticResponseMesoporous2015} further enhancing their potential for responsive nanoacoustic applications. Another key factor is the pore size distribution, which plays a crucial role in adsorption capacity, as capillary condensation exhibits a strong nonlinear response to vapor pressure variations.\cite{gimenezPreparationMesoporousSilica2020,cardozodeoliveiraDesignCosteffectiveEnvironmentresponsive2023}

Characterizing the mechanical properties of MTFs is crucial for designing optimized devices. Nowadays, several techniques have been adapted to overcome limitations of nanoindentation due to the presence of the substrate.\cite{lionelloStructuralMechanicalEvolution2017a,lionelloMechanicalPropertiesOrdered2022,boissierePorosityMechanicalProperties2005} A novel technique using plasmonic nanobars have been recently employed to characterize the mechanical properties of MTFs.\cite{boggianoOpticalReadoutMechanical2023} The direct access to acoustic resonances within the mesoporous materials is another strategy to characterize MTFs.

Previous studies have demonstrated the confinement of GHz acoustic phonons in MTFs based on silicon dioxide (SiO$_{2}$) and titanium dioxide (TiO$_{2}$).\cite{abdalaMesoporousThinFilms2020,cardozodeoliveiraProbingGigahertzCoherent2023} However, these investigations focused on structures where the mesoporous layer was embedded between the substrate and a metallic optoacoustic transducer. This configuration presents two major limitations: (1) the metallic capping layer obstructs liquid and vapor infiltration into the pores, restricting environmental responsiveness, and (2) the coherent phonon detection, done with ultrafast techniques, requires access to advanced interferometric measurement setups (e.g., Sagnac configuration).\cite{perrinInterferometricDetectionPicosecond1999}

Here, we explore open-cavity acoustic resonances in structures where the SiO$_{2}$ MTFs form the topmost layer, directly exposed to the environment. We show that a conventional transient reflectivity setup is enough to detect the acoustic resonances within the mesoporous materials. We demonstrate the environmental responsiveness of these acoustic resonators by controlling the environment relative humidity (RH), i.e., water vapor pressure. Additionally, we investigate two groups of samples with different pore diameters and three distinct MTF thicknesses. The results indicate a clear dependence of the acoustic resonance on the film thickness, while the two groups of pore sizes exhibit similar responses. This suggests that the acoustic resonances are primarily governed by the material properties and layer thickness rather than the pore configuration. Furthermore, numerical simulations exhibit a good agreement with the experimental results, further validating our approach.

Overall, our work offers a new platform for developing cost-effective tunable nanoacoustic devices. This open-cavity design will enable the study of acoustic responsivity to environmental conditions.

\section{Experimental details}

\begin{figure}
	\centering
	\includegraphics[width=\linewidth]{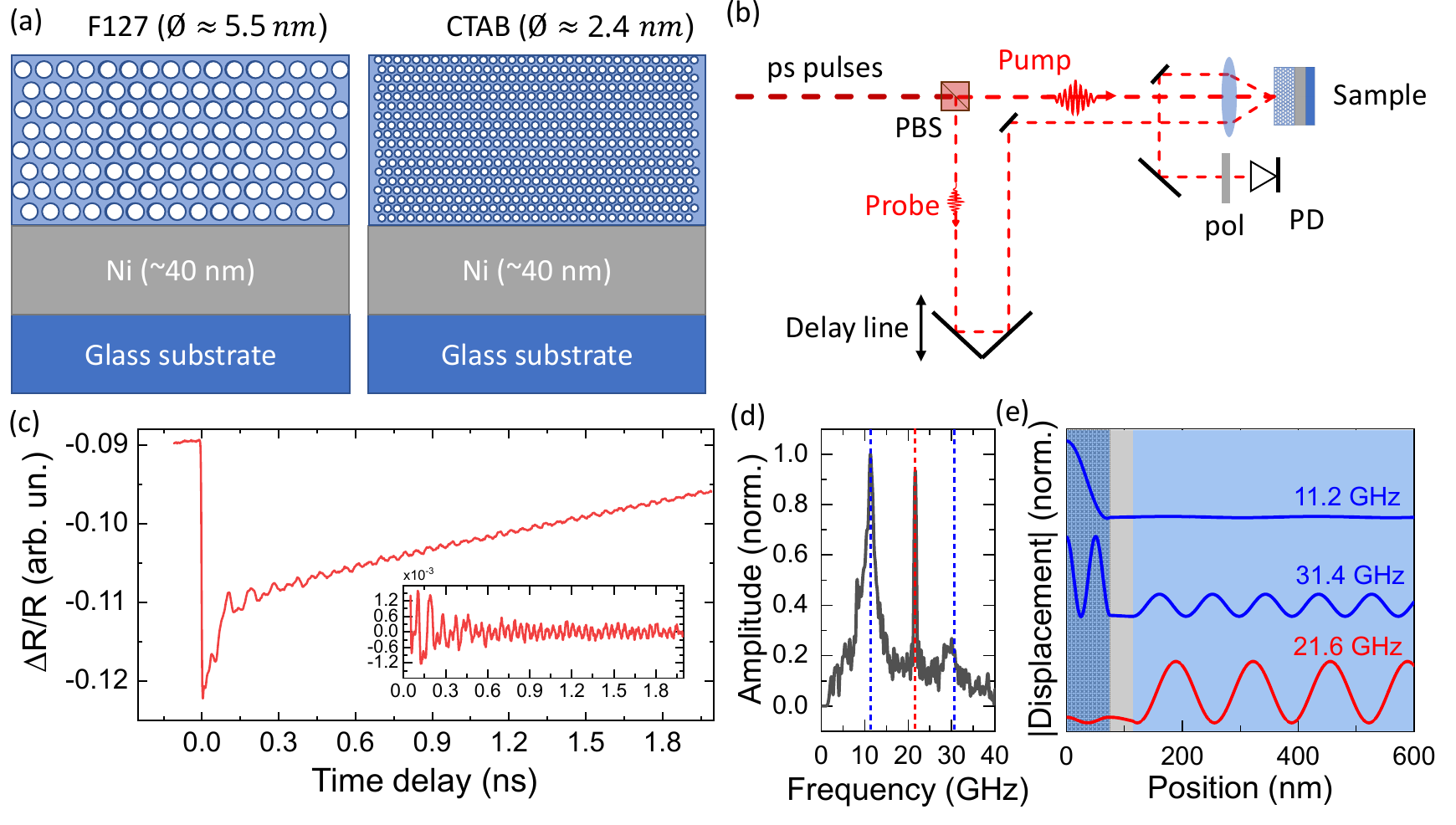}
	\caption{(a) Schematic representation of the sample designs. The left and right samples represent the F127 (larger pores) and CTAB (smaller pores) groups, respectively. (b) Experimental setup scheme of the reflectometric pump-probe setup. (c) Transient reflectivity time trace of the sample F127-1. Inset displays the processed time trace with a polynomial fit. (d) FFT spectrum of the timetrace displayed in (c). (e) Displacement field of the modes at 11.2, 21.6 and 31.4 GHz. Blue and red lines correspond to the modes within the MTF and the Brillouin oscillations, respectively, and are also represented as dashed lines in panel (d).}
	\label{fig:sample_setup_timetrace}
\end{figure}

We studied samples composed of glass substrates covered by 40-nm-thick Nickel layers, followed by SiO$_{2}$ mesoporous thin films serving as the resonant open-cavity, as illustrated in Figure~\ref{fig:sample_setup_timetrace}(a). Two different templates were used to provide distinct pore diameters: cetyltrimethylammonium bromide (CTAB) and Pluronic F127. The F127 and CTAB compounds yield mesopores with average diameters of $\sim$5.5~nm and $\sim$2.4~nm, respectively, according to ellipsoporosimetry analysis (see Supplementary Information).\cite{boissierePorosityMechanicalProperties2005,gazoniDesignedNanoparticleMesoporous2017} The mesoporous thin film thicknesses ranged from 75~nm to 150~nm, depending on the dip-coating withdrawal speed and the viscosity of the solutions (see Supplementary Information for further details). The samples are labeled F127-X and CTAB-X, where X = 1, 2, 3 corresponds to different film thicknesses (refer to Table~\ref{table:samples}).

\begin{table}[hb]
	\centering
	\caption{Thickness values of the mesoporous layers for the samples from group F127 and CTAB, and the withdrawal speed used during the dip coating.}
	\begin{tabularx}{\linewidth}{cXc}
		\toprule
		Sample & \makecell[c]{Thickness [nm]} & \makecell[c]{Withdrawal \\ speed [mm s$^{-1}$]} \\ 
		\midrule
		F127-1 & \makecell[c]{75} 		 &	1.8  \\
		F127-2 & \makecell[c]{115} 		 & 	3.4  \\
		F127-3 & \makecell[c]{150} & 	5.2  \\
		\midrule
		CTAB-1 & \makecell[c]{78}  		 & 3.2  \\
		CTAB-2 & \makecell[c]{112} 		 & 5.7  \\
		CTAB-3 & \makecell[c]{150} 		 & 11.7  \\
		\midrule
	\end{tabularx}
	\label{table:samples}
\end{table}

The nanoacoustic response characterization is done by employing a coherent phonon generation and detection experiment with a reflectivity pump-probe setup, illustrated in Figure~\ref{fig:sample_setup_timetrace}(b).\cite{thomsenCoherentPhononGeneration1984,thomsenSurfaceGenerationDetection1986} Experimental details about the technique can be found in the Experimental Section. A reflectivity timetrace is displayed in Figure~\ref{fig:sample_setup_timetrace}(c). At time delay $t=0$~ps, when pump and probe pulses simultaneously reach the sample, an ultrafast reflectivity change occurs due to electronic excitation, followed by a gradual decay over more than 2~ns as a result of thermal relaxation. Oscillations in the reflectivity are also superposed to this decay, associated with the photoinduced coherent acoustic phonons. A polynomial fit is applied to the time trace to remove the slow relaxation component, and the residual is displayed in the inset of Figure~\ref{fig:sample_setup_timetrace}(c). Finally, a fast Fourier transform (FFT) of this processed signal results in the phononic spectrum (see Figure~\ref{fig:sample_setup_timetrace}(d)). 

\section{Results and Discussion}

\begin{figure}
	\centering
	\includegraphics[width=0.7\linewidth]{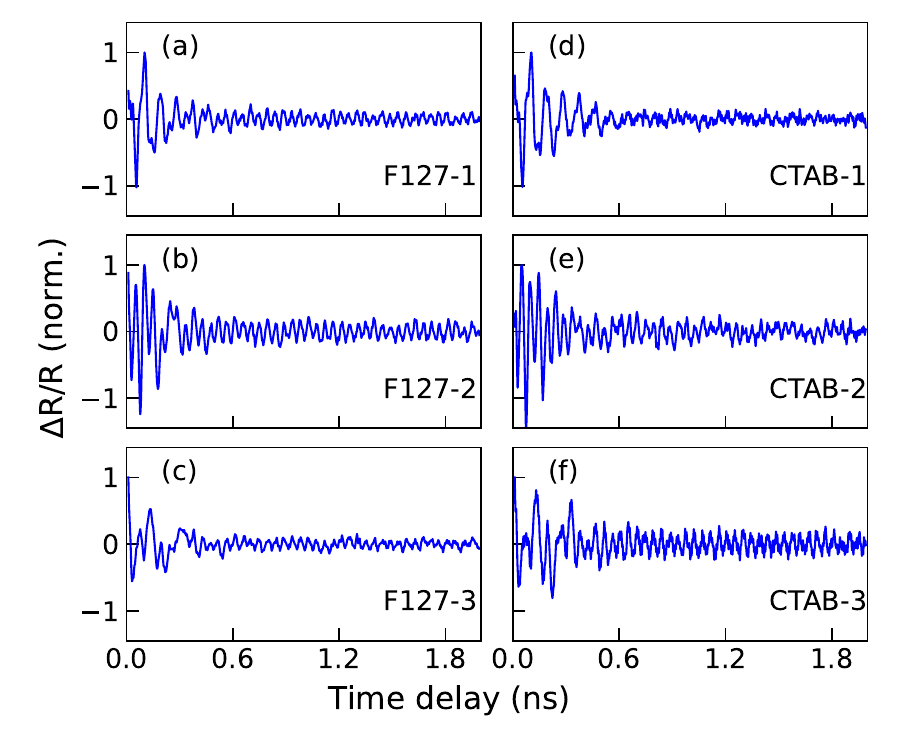}
	\caption{Transient reflectivity time trace of the samples from the F127 [panels (a)-(c)] and CTAB [panels (d)-(f)] groups. See Table \ref{table:samples} for more information on the samples.}
	\label{fig:timetraces}
\end{figure}

Figure~\ref{fig:timetraces} displays the reflectivity time traces for all measured samples over a time window of two nanoseconds. For both sample groups, F127 (Figure~\ref{fig:timetraces}(a)-(c)) and CTAB (Figure~\ref{fig:timetraces}(d)-(f)), two major oscillatory contributions are observed: strong oscillations that decay within the first 0.6~ns and weaker, long-lived oscillations spanning over the entire measured time window. The former oscillations are directly linked to resonances within the MTFs, whose periods exhibit a clear dependence on layer thickness. The latter contribution, on the other hand, corresponds to the so called Brillouin oscillations, which arise from the photoinduced strain pulse propagating through the substrate,\cite{sandeepInSituImagingLightInduced2022,gusevAdvancesApplicationsTimedomain2018} and are dependent on the probe wavelength, sound velocity and refractive index on the material. Thus, the Brillouin oscillations feature the same frequency across all samples.

\begin{figure}
	\centering
	\includegraphics[width=0.7\linewidth]{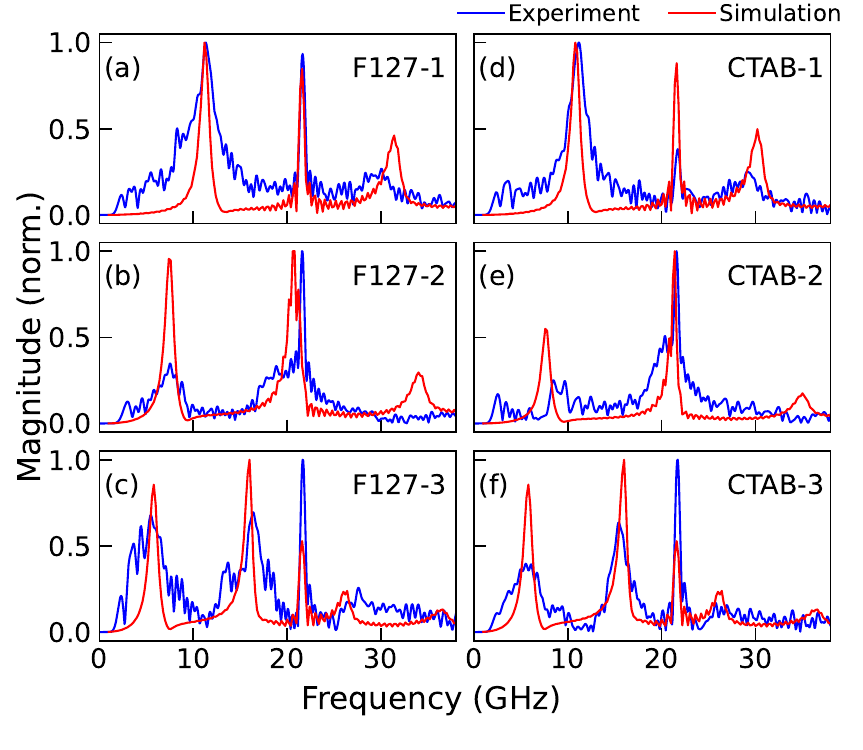}
	\caption{Coherent phonon generation-detection spectra of the samples from the F127 [panels (a)-(c)] and CTAB [panels (d)-(f)] groups. Blue lines correspond to the experimental Fourier transform spectra from the respective time traces displayed in Figure~\ref{fig:timetraces}. Red lines are associated with the numerical simulation results.}
	\label{fig:expxsim}
\end{figure}

The experimental FFT results are shown in Figure~\ref{fig:expxsim}(a)-(c) for the F127 group and Figure~\ref{fig:expxsim}(d)-(f) for the CTAB group. For instance, the sample F127-1 [Figure~\ref{fig:expxsim}(a)] features three distinct peaks: two at 11.2 GHz and 31.4 GHz corresponding to the fundamental and first harmonic acoustic resonances from the MTF, and a peak at 21.6 GHz attributed to the Brillouin oscillations on the glass substrate. By analyzing the resonant frequency peaks of thicker MTF samples [Figure~\ref{fig:expxsim}(b) and (c)], noticeable redshift can be observed, consistent with expectations for acoustic confinement. A similar trend is observed for the CTAB group [Figure~\ref{fig:expxsim}(d)-(f)]. The Brillouin peak remains fixed across all the samples. Variations in the relative amplitude of the MTF resonances and the Brillouin peak can be attributed to factors such as fluctuations in pump power, sample inhomogeneities, light scattering, and minor deviations in optical focusing.

To further interpret the experimental results and distinguish the contributions of each layer to phonon detection, we performed numerical simulations based on the transfer matrix method.\cite{cardozodeoliveiraDesignCosteffectiveEnvironmentresponsive2023,lanzillotti-kimuraTheoryCoherentGeneration2011,matsudaReflectionTransmissionLight2002} These simulations accounted for both the optical electric field distribution and the acoustic displacement profile, incorporating the photoelastic model to describe the generation and detection of acoustic phonons. The acoustic generation and detection spectra were calculated using an overlap integral between the strain field (i.e., the displacement derivative), the electric field, and a spatially dependent transduction constant. For the generation, this constant represents, in our case, the thermoelastic capacity of the material $K$, whereas for the detection process, the constant corresponds to the photoelastic coefficient $p$. The simulation was performed following the approach detailed in \cite{lanzillotti-kimuraTheoryCoherentGeneration2011,cardozodeoliveiraDesignCosteffectiveEnvironmentresponsive2023}. In our model, coherent phonon generation is assumed to be entirely limited to the nickel transducer layer, where the strong thermoelastic effect dominates, making the transduction term $K$ nonzero only in this region. Phonon detection, however, occurs across all layers due to strain-induced modulation of the refractive index, resulting in reflectivity changes throughout the sample. Therefore, nonzero photoelastic terms in the detection process $p$ are considered for the glass substrate and the MTF. We also assume the mesoporous materials as solid layers with effective optical and elastic properties, as well as a penetration depth of 90~nm to account for acoustic attenuation.\cite{abdalaMesoporousThinFilms2020,cardozodeoliveiraProbingGigahertzCoherent2023} The layer thicknesses and material parameters used in the simulations are provided in Table~\ref{table:samples} and \ref{table:parameters}, respectively. 

\begin{table}[hb]
	\centering
	\caption{Optical and elastic properties of mesoporous SiO$_{2}$ (mSiO$_{2}$), Ni and glass used in the numerical simulations. $n$, $v$, $\rho$, $K$~(gen.) and $p$~(det.) correspond to refractive index, sound velocity, mass density generation transduction term and photoelastic term used to calculate the acoustic phonon generation and detection processes, respectively.}
	\begin{tabularx}{\linewidth}{XXXXXX}
		\toprule
		Material & $n$ & $v$ [m s$^{-1}$] & $\rho$ [g cm$^{-3}$] & $K$ (gen.) & $p$ (det.)\\ 
		\midrule
		mSiO$_{2}$\cite{abdalaMesoporousThinFilms2020}	& 1.323			& 3156.7	& 1.32	& 0 & 0.02   \\
		Ni\cite{cardozodeoliveiraDesignCosteffectiveEnvironmentresponsive2023,cardozodeoliveiraProbingGigahertzCoherent2023} 			& 2.22+4.89i	& 5580		& 8.908	& 1 & 1      \\
		Glass		& 1.50			& 5750		& 2.2	& 0 & 0.1    \\
		\midrule
	\end{tabularx}
	\label{table:parameters}
\end{table}

The calculated displacement profiles for the three main peaks of the sample F127-1, depicted in Figure~\ref{fig:sample_setup_timetrace}(e), confirm that the resonances at $\sim$11 GHz and $\sim$30 GHz correspond to modes within the MTF, whereas the peak at $\sim$22 GHz is related with the Brillouin oscillations on the substrate. 

The resulting spectra, which represent the product of the phonon generation and detection processes, are presented in Figure~\ref{fig:expxsim}. Notably, the simulations show excellent agreement with experimental data, with well matched frequencies for the observed peaks. The good agreement between simulation and experiment further supports our interpretation of the observed resonances.

To investigate the acoustic response of MTFs to humidity variations, pump-probe experiments were performed under controlled relative humidity conditions on the CTAB-3 sample. The RH was adjusted using a commercial humidity controller (MHG32-TC) within a sealed chamber, with measurements acquired at RH levels of 4\%, 14\%, 24\%, 34\%, 44\%, 64\% and 84\%.

\begin{figure}
	\centering
	\includegraphics[width=0.7\linewidth]{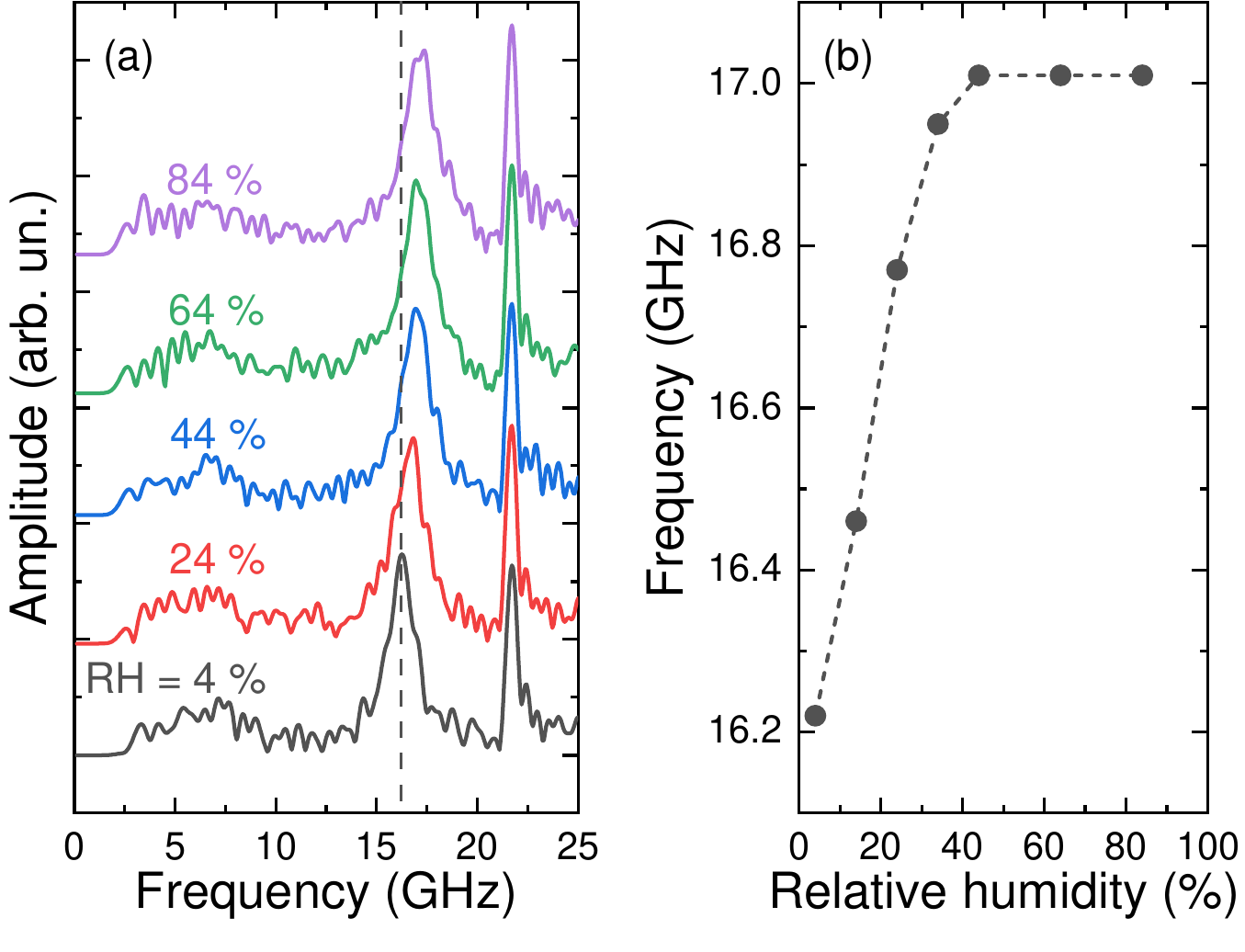}
	\caption{(a) FFT spectra of the CTAB-3 MTF under different relative humidity levels. (b) Dependence of the first harmonic frequency on relative humidity.}
	\label{fig:humidity}
\end{figure}

Figure~\ref{fig:humidity}(a) presents the corresponding FFT spectra for selected humidity conditions. At low humidity (RH = 4\%), the spectrum exhibits acoustic modes comparable to the measurements displayed in Figure~\ref{fig:expxsim}(f), with minor frequency differences attributed to inhomogeneities in the film thickness across the sample. As humidity increases, the general spectral profile remains consistent, but a frequency shift is observed in the first harmonic peak. The broad fundamental mode does not display a clearly resolved shift due to its larger linewidth.

This frequency tuning behavior is summarized in Figure~\ref{fig:humidity}(b), which shows an increase in the first harmonic frequency with increasing RH, and reaching a saturation level near 40\% RH. The total frequency shift reaches approximately 0.8 GHz, demonstrating a strong sensitivity of the resonator to environmental humidity. This behavior highlights the potential of mesoporous-based nanophononic resonators as tunable and responsive devices in the GHz frequency range.

Nevertheless, reproducibility remains an open question, as humidity cycles could lead to structural changes, e.g., pore clogging or nickel hydroxide formation. These issues may be mitigated by modifying the fabrication process—for instance, performing calcination under an inert atmosphere (e.g., nitrogen) to prevent nickel oxidation. Despite these challenges, the results represent a significant step toward the development of environment-responsive and tunable nanophononic platforms.

\section{Conclusion}

In summary, we have demonstrated the tunable confinement of nanoacoustic resonances in a novel design of SiO$_{2}$ MTFs, where the mesopores are exposed to the environment. Unlike previous designs requiring complex interferometric detection, this configuration enables the use of conventional transient reflectivity pump-probe setups to effectively detect acoustic resonances. This design allowed us to showcase the responsivity of these resonators towards the environment (in this case, water vapor pressure), demonstrating, for the first time, the possibility of externally reconfigurable hypersound devices. Analysis of layer thickness and pore sizes, combined with numerical simulations, further supports the validity of our approach. 

Previous studies have also proposed an alternative design that incorporates an acoustic distributed Bragg reflector (DBR) beneath the acousto-optical transducer.\cite{cardozodeoliveiraDesignCosteffectiveEnvironmentresponsive2023} This would minimize phonon leakage into the substrate,\cite{lanzillotti-kimuraAcousticPhononNanowave2007} thereby enhancing the quality factor of the resonator. These advances provide a novel tool for studying the structural properties of the MTFs. Additionally, controlling the pore size distribution opens new possibilities for sensing applications and exploring fundamental nanoscale phenomena such as the dynamics of condensation of liquids in nanostructures. This framework pushes the boundary for tunable and responsive nanophononic applications.

\section{Experimental Section}
\subsubsection*{Sample fabrication}

The 40-nm-thick Nickel layers were deposited on glass substrates via electron beam evaporation. The mesoporous thin films were then deposited using dip-coating from absolute ethanol-water solutions. To obtain different pore size distributions, two distinct solutions were prepared. The first solution consists of a Pluronic F127 copolymer as the mesopore template, with preparation details described in~\cite{abdalaMesoporousThinFilms2020}. The second solution is based on cetyltrimethylammonium bromide (CTAB), a quaternary ammonium compound, as the mesopore template, and tetraethyl orthosilicate (TEOS), from Sigma-Aldrich, as the oxide precursor. To optimize the chemical reactivity of the silica precursor, a pre-hydrolysis step was performed, consisting of 1 hour agitation at 50~\textdegree C for the TEOS/EtOH/H$_{2}$O/CTAB solution with a 1:40:5:0.1 molar ratio. The mesoporous thin film thicknesses ranged from 75~nm to 150~nm, depending on the dip-coating withdrawal speed and the viscosity of the solutions (see Supplementary Information for further details). The mesoporous films are formed by the evaporation-induced self-assembly approach. After the dip-coating, both solutions went through a post-treatment procedure consisting of an aging and thermal stabilization process. This consisted of three 30-minute steps: first in a chamber at room temperature and 50\% relative humidity, followed by thermal treatments at 60~°C and 130~°C. This treatment allows to obtain ordered mesoporosity and to consolidate the mesoporous film structure.\cite{abdalaMesoporousThinFilms2020,gazoniDesignedNanoparticleMesoporous2017} Finally, a calcination step of 2~h at 350~°C with a controlled heating ramp of 1~°C/min is done to eliminate the pore template.

\subsubsection*{Coherent phonon generation and detection}

The impulsive generation of acoustic phonons in the Ni transducer via thermoelastic effect\cite{ruelloPhysicalMechanismsCoherent2015} is done with a pulsed laser delivering 2.5~ps pulses at 800~nm, with average powers of 20-30~mW and at a repetition rate of 80~MHz. The pump beam passes through an acousto-optical modulator at 800 kHz for a lock-in detection. The ultrafast strain profile is developed in the structure, altering the optical refractive index of the materials. A time-delayed probe pulse, with the same specifications as the pump but with an average power of 5~mW, was used to monitor these refractive index changes, resulting in a reflectivity modulation. The signal is demodulated by a lock-in for improving the signal-to-noise ratio. By scanning the relative time delay between the pump and probe, the reflectivity time trace is reconstructed.

\medskip

\section*{Acknowledgements}
The authors acknowledge funding by the French RENATECH network and the CNRS International Research Project Phenomenas. N.D.L.-K., C.X. and E.R.C. de O. acknowledge funding from European Research Council Consolidator Grant No. 101045089 (T-Recs). G.J.A.A.S.-I. acknowledges project AFOSR (award No. FA9550-24-1-0209), ANPCyT for projects PICT 2017-4651, PICT-2018-04236, and PICT 2020-03130, and NANOQUIMISENS Red Federal de Alto Impacto \#87.

\end{document}

\renewcommand{\thefigure}{S\arabic{figure}}

\title{Supplemental Material for Towards Environmentally Responsive Hypersound Materials}

\author[1]{Edson R. Cardozo de Oliveira}
\author[2]{Gastón Grosman}
\author[1]{Chushuang Xiang}
\author[2]{Michael Zuarez-Chamba}
\author[2]{Priscila Vensaus}
\author[1]{Abdelmounaim Harouri}
\author[3]{Cédric Boissiere}
\author[2]{Galo J. A. A. Soler-Illia\footnote{gsoler-illia@unsam.edu.ar}}
\author[1]{Norberto Daniel Lanzillotti-Kimura\footnote{daniel.kimura@cnrs.fr}}

\affil[1]{Université Paris-Saclay, C.N.R.S., Centre de Nanosciences et de Nanotechnologies (C2N),  10 Boulevard Thomas Gobert, 91120 Palaiseau, France}

\affil[2]{Instituto de Nanosistemas, Escuela de Bio y Nanotecnología, Universidad Nacional de San Martín-CONICET, Av. 25 de Mayo 1169, San Martín, Buenos Aires, Argentina}

\affil[3]{Laboratoire de Chimie de la Matière Condensée, Université Pierre et Marie Curie, 4 Place Jussieu, 75252 Paris, Cedex 5, France}

\date{}

\maketitle

\section{Determination of pore size distribution of F127 and CTAB templates}

Atmospheric ellipsometry porosimetry (EPA) measurements were performed on mesoporous thin films (MTFs) deposited on silicon substrate to estimate the pore size distribution of the templates CTAB and F127. Figures S1(a) and (b) display the isotherms of water infiltration into the pores under adsorption and desorption cycles as a function of the relative humidity (RH), for F127 and CTAB, respectively. In the adsorption curve the F127-based MTFs show a wetting threshold at $\sim$70\% of RH, and a reduction threshold in the water content at $\sim$40\%, resulting in a pronounced hysteresis curve. The film with CTAB template, on the other hand, undergoes a steep transition in water content at $\sim$50\% of RH for both the adsorption and desorption curves. 

\begin{figure}[h!]
    \begin{center}
        \includegraphics[width=0.8\textwidth]{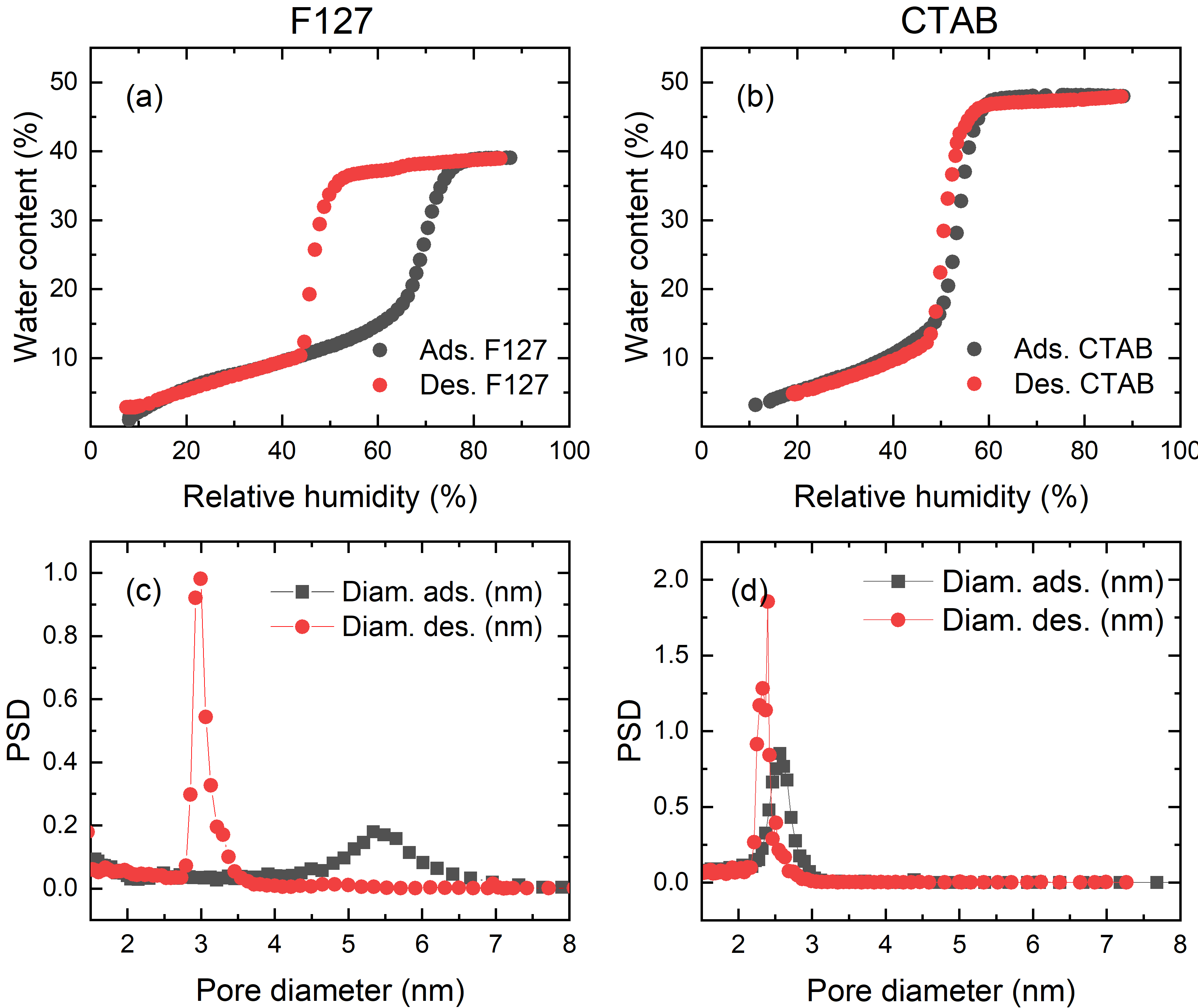}
    \end{center}
    \caption{Adsorption-desorption isotherms of water content versus relative humidity for stabilized (a) F127 and (b) CTAB films. (c) Pore size distribution (PSD) of the (c) F127 and (d) CTAB films. }
    \label{figs1}
\end{figure}

From these measurements, the pore size distribution (PSD) then can be obtained by implementing the isotropic inorganic pore contraction (IIC) model.\cite{boissierePorosityMechanicalProperties2005} The PSD results for F127 and CTAB templates are displayed in Figs. S1(c) and (d), respectively. Under adsorption cycle, the PSD of the F127 film is ~5.5 nm, whereas the CTAB film displays a pore diameter of around 2.6 nm. Under desorption cycles, there is a known contraction of the films, leading to 3 nm pore diameter for F127, and 2.3 nm for CTAB. The porosity obtained for F127 and CTAB are, respectively, $\sim$38\% and $\sim$48\%, extracted from the corresponding isotherms. EPA analyses were performed in following conditions:

\begin{itemize}\setlength\itemsep{0pt}
	\item Temperature of atmosphere: 25\,°C.
	\item Substrate: Silicon wafer, (100) orientation.
	\item Three adsorption/desorption cycles were made before data were taken (10--90\% relative humidity in 5 minutes, and 90--10\% relative humidity in 5 minutes).
	\item Relative humidity ramp: 0--100\% in 20 minutes and 100--0\% in 20 minutes.
	\item Ellipsometer: M2000 from Woolam.
	\item Model: Silicon/native oxide (2 nm)/Cauchy model with $A$, $B$, and thickness parameters being fitted.
	\item Fitting performed in the 370--1390 nm wavelength range.
	\item Refractive index value given at 700 nm.
\end{itemize}

\pagebreak

\section{Determination of pore size distribution of F127 and CTAB templates}

To estimate the desired thickness of the mesoporous thin films, calibration curves were established by correlating film thickness with the dip-coating withdrawal speed for each template. For the CTAB template, three reference samples were prepared by dip-coating silicon substrates at withdrawal speeds of 1 mm/s, 2 mm/s and 8 mm/s. After deposition, the samples were submitted to an aging process consisting of 5 minutes at 40\% relative humidity, followed by 10 minutes at 60 °C. X-Ray reflectometry (XRR) measurements, shown in Fig. S2(a), were then performed to determine the film thicknesses, extracted from the interference fringes in the reflectivity data. XRR measurements were repeated after calcination at 350 °C for 2 hours to remove the pore template, allowing quantification of the film contraction, which was found to be in average around 15\%. A calibration curve was then constructed by fitting the measured thicknesses as a function of the withdrawal speed raised to the 2/3 power, a standard scaling behavior for mesoporous thin films.\cite{landauDynamicsCurvedFronts1988,kuszRoleMolecularInteractions2024} This curve was subsequently used to estimate target film thicknesses.

\begin{figure}[h!]
    \begin{center}
        \includegraphics[width=0.8\textwidth]{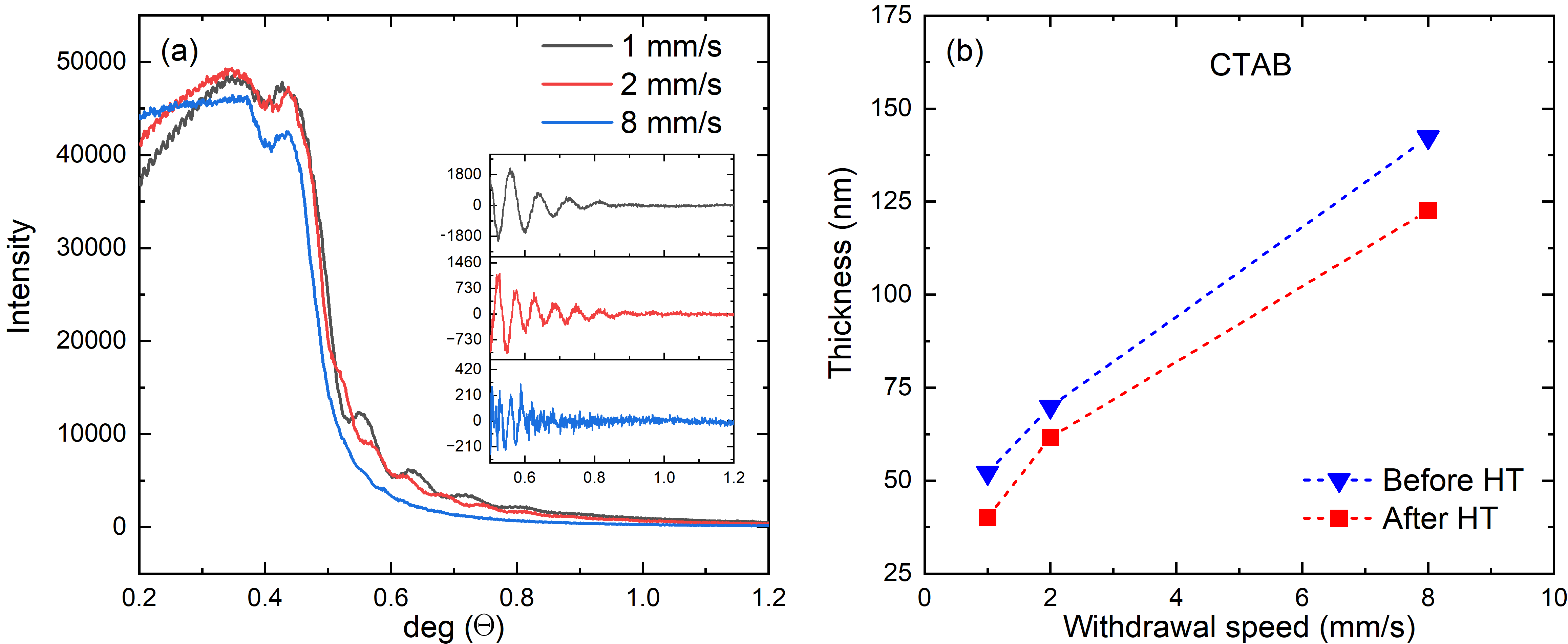}
    \end{center}
    \caption{(a) X-ray reflectometry (XRR) of mesoporous thin films prepared with the CTAB template at withdrawal speeds of 1, 2, and 8~mm/s, represented by black, red, and blue lines, respectively. The inset highlights the oscillation patterns for each speed, obtained by fitting the decay. (b) Film thickness versus withdrawal speed, measured before (blue triangles) and after (red squares) heat treatment at 350~°C for 2 hours. An average thickness reduction of $\sim$15\% was observed.}
    \label{figs2}
\end{figure}

For the F127 template, the calibration curve was estimated based on the data reported in the PhD thesis of M. C. Fuertes.\cite{fuertesMaterialesFuncionalesMultiescala2009} To account for differences in solution viscosity and concentration it was fabricated one sample in the same conditions as the literature, and a scaling factor of 0.53 was obtained. Both calibration curves – for the CTAB and F127 templates – are presented in Fig. S3.

\begin{figure}[h!]
	\begin{center}
		\includegraphics[width=0.5\textwidth]{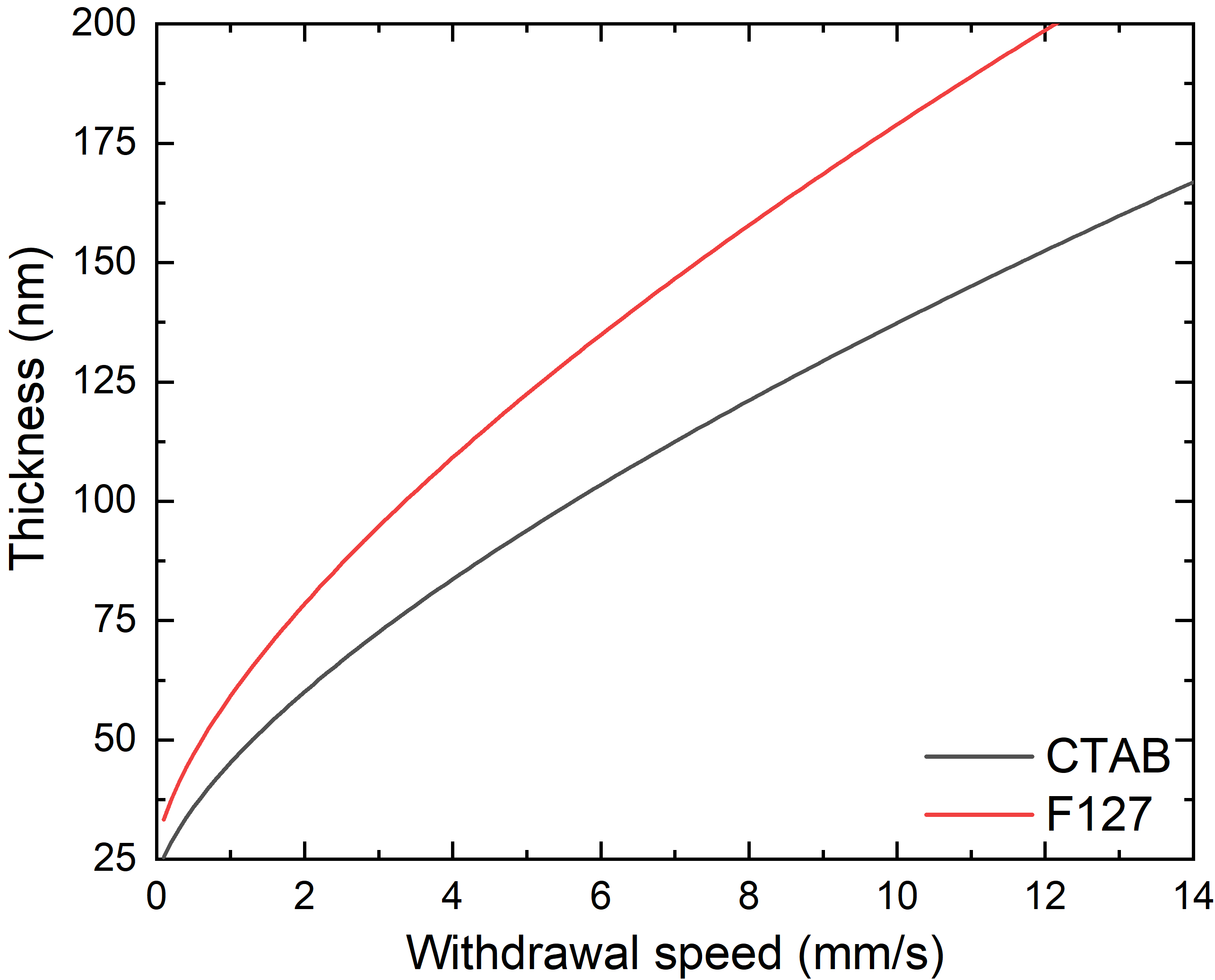}
	\end{center}
	\caption{Calibration curves of film thickness as a function of dip-coating withdrawal speed for mesoporous films prepared using CTAB (black) and F127 (red) templates. The data for the CTAB template were obtained experimentally, while the F127 curve is based on previously reported results. Both curves follow the characteristic $v^{2/3}$ dependence typical of mesoporous film deposition.}
	\label{figs3}
\end{figure}

\pagebreak

\section*{Acknowledgements}
The authors acknowledge funding by the French RENATECH network and the CNRS International Research Project Phenomenas. N.D.L.-K., C.X. and E.R.C. de O. acknowledge funding from European Research Council Consolidator Grant No. 101045089 (T-Recs). G.J.A.A.S.-I. acknowledges project AFOSR (award No. FA9550-24-1-0209), ANPCyT for projects PICT 2017-4651, PICT-2018-04236, and PICT 2020-03130, and NANOQUIMISENS Red Federal de Alto Impacto \#87.